\documentclass[12pt]{article}
\usepackage{amssymb,amsmath}
\begin{document}

\sloppy
\title
{\Large  Gravity from the viewpoint of theory of sources  }

\author
{
A.I.Nikishov
\thanks
{E-mail: nikishov@lpi.ru}
\\
{\small \phantom{uuu}}
\\
{\it {\small} I.E.Tamm Department of Theoretical Physics,}
\\
{\it {\small} P.N.Lebedev Physical Institute, Moscow, Russia}
\\
}
%
\maketitle
\begin{abstract}

I examine the $G^2$-approximation of Schwarzschild solution from the viewpoint of theory of sources. The method suggests the following definition of the privileged coordinate system: it is a system in which in each approximation
the gauge degrees of freedom are put to zero, i.e. the metric is formed solely by sources. I calculate the metric in this system. In $G^2$-approximation the exterior metric has the term which is of the form of a gauge function. Considering it as such I have the agreement with Schwarzschild metric. But I cannot consider it as a gauge function
because it is generated by sources. It should be observable. It is proportional to the radius of matter ball 
and seems 
violate the Birckhoff theorem.

\end{abstract}
\section{Introduction}
From the viewpoint of theory of sources the metric is determined by sources.
In this paper I assume that the sources are given by Einstein equation written
in field theoretical form by Weinberg, see \S 6 in Ch.7 in [1].

It is instructive to examine the $G^2$-approximation of Schwarzschild solution in
order to obtain an experience in tackling problems encountered in investigating a
modified 3-graviton vertex, see for example [3 ].

In general relativity the metric is obtained from Einstein's gravitational equation which is a differential one.
In theory of sources the metric is defined (via an integral equation) by sources not only in the
neighbourhood of the point under consideration but also at some distance from it.
So in general one may expect deviations from general relativity and this seems
really happened: the calculated  metric of the external solution feels the presence of ball of
matter. 
 The following  notation is used
$$
 g_{ik}=\eta_{ik}+h^{(1)}_{ik}+ h^{(2)}_{ik}+\cdots,\quad h^{(n)}_{ik}\propto
G^n,\quad\eta_{ik}={\rm diag}(-1,1,1,1),\quad \frac{\partial h}{\partial x^{\alpha}}=h_{,\alpha}.                                           \eqno(1.1)
$$
The peculiarity of the problem is the appearance of two solutions in $G^2$-
approximation. In linear approximation the solution
$h^{(1)}_{ik}(m',r)=-2\phi\delta_{ik}$, ($\phi$ is the Newtonian potential)
depends on initial mass
$m'=\frac{4}{3}\pi\mu b^3$, where $b$ is the radius of the ball of matter and
$\mu$ is density. But already in $G^2$-approximation  in
external solution appear terms, which  have the same $\frac1r$ form as in linear solution. Joint to the
latter they dress mass $m'$ to observable mass $m$.
It turns out that $m=m'(1+\frac{3mG}{b})$.

The solutions $h^{(2)}_{ik}$ obtained by perturbation theory I denote as
$h^{(2)}_{ik}(m',r)$ although they may depend also on $x_{\alpha}x_{\beta}$.
The $m'$
in the argument indicate only the origin of solution, not a functional dependence upon it. In the expression for
this solution (i.e. in the r.h.s. of definition) it does not matter whether we
write $m$ or $m'$.

The same result we get in the following way:  inserting $m=m'(1+\frac{3mG}{b})$
in linear solution $h^{(1)}_{ik}(m',r)$ we get $h^{(1)}_{ik}(m,r)$ plus terms
proportional to $G^2$. Joining these terms to $h^{(2)}_{ik}(m',r)$  just cancel the mass dressing terms there and we get by
definition $h^{(2)}_{ik}(m,r)$. Thus,
$$
h^{(1)}_{ik}(m',r)+h^{(2)}_{ik}(m',r)=h^{(1)}_{ik}(m,r)+h^{(2)}_{ik}(m,r).         \eqno(1.2)
$$

In Section 1 I find the interior metrics in harmonic and isotropic coordinate systems: first from exact solution in standard system, then  from gravitation equation by perturbation theory and finally from theory of sources. In Section 2 I obtain the  exterior solution in the privileged coordinate system. In Appendix I collect formulas used in calculating the metrics.

\section{Interior solutions}
{\bf Obtaining harmonic coordinate solution from exact solution in standard coordinates}\\
The interior metric in standard coordinates has the form, see Synge [4], eq.(7.183)
$$
ds^2=\frac{dr^2}{1-qr^2}+r^2(d\theta^2+\sin^2\theta d\varphi^2)-\left(\frac{3\sqrt{1-qa^2}  -\sqrt{1-qr^2}}{2} \right)^2dt^2, \quad q=\frac{2mG}{a^2},                                         \eqno(2.1)                           $$
$a$ is the radius of ball of liquid. The transition from standard coordinate $r$ to harmonic one, $R$,  in exterior region is
given by the relation $r=R+mG$. To find its counterpart for interior region, we rewrite this relation in the form
$$
r=R(1-\phi(m,R)),\quad  \phi(m,R)=-\frac{mG}{R}.                                                                                                                                                           \eqno(2.2)
$$
Now it is enough to consider $\phi(m,R)$ as Newtonian potential inside the ball:
$$
\left.\phi(m,R)\right|_{R<b}=\phi(m,b.R)=\frac{mG}{2b}(\frac{R^2}{b^2}-3).                                                                                                                                  \eqno(2.3)
$$
Then using the relations
$$
dr=[ 1+\frac{3mG}{2b}\left(1-\frac{R^2}{b^2}\right)]dR,\quad RdR=X_{\alpha}dX_{\alpha},\quad (dR)^2=\frac{X_{\alpha}X_{\beta}dX_{\alpha}dX_{\beta}}{R^2}, \quad \alpha,\beta=1,2,3,    \eqno(2.4)                            $$
we obtain from (2.1)
$$
g_{\alpha\beta}(\vec X)=1-2\phi(m,b.R)+h^{(2)harm}_{\alpha\beta}(m,R), \eqno(2.5)
$$
$$
h_{\alpha\beta}^{(2)harm}(m,R)=
\frac{m^2G^2}{b^2}[\delta_{\alpha\beta}(\frac{9}{4}-\frac{3R^2}{2b^2}+\frac{R^4}{4b^4})+(3-
\frac{2R^2}{b^2})\frac{X_{\alpha}X_{\beta}}{b^2})].                                     \eqno(2.6)         
$$

For $g_{00}$ we obtain from (2.1)
$$
g_{00}=-[1+2\phi(m,a,r)]+\frac{m^2G^2}{b^2}\left(-\frac34+\frac{3R^2}{2b^2}-\frac{3R^4}{4b^4}\right).                           \eqno(2.7)
$$
Expressing in $\phi(m,a,r)$ the parameter $a=b+mG$ and $r$ (see the first equation in (2.2) and equation (2.3)) in terms of $b$ and $R$ we find
$$
2\phi(m,a,r)=\frac{mG}{a}(\frac{r^2}{a^2}-3)=2\phi(m,b,R)+\frac{m^2G^2}{b^2}[3-\frac{R^4}{b^4}].\eqno(2.8)
$$
Then from (2.7) and (2.8) we get
$$
g_{00}=-[1+2\phi(m,b,R)]+h^{(2)}_{00}(m,R),\quad
h^{(2)}_{00}(m,R)=\frac{m^2G^2}{b^2}\left(-\frac{15}{4}+\frac{3R^2}{2b^2}+\frac{R^4}{4b^4}\right). \eqno(2.9)
$$
In isotropic coordinates in the considered approximation the radius of the ball of matter remains the same $b$.

  Here we make farewell to standard Schwarzschild coordinates and denote by $\vec x$ what was up to now $\vec X$. So from now on we deal only with coordinates  $\vec x$:
in isotropic coordinates $g_{00}$ is the same as in harmonic one because   the sources, defined by the linear approximation, are the same. For the same reason $h_{\alpha\beta}^{iso}$ may differs from $h_{\alpha\beta}^{(2)harm}$ only by gauge. We can obtain $h^{iso}_{ik}$ from Rosen paper [5]. In our approximation
$$
h^{(2)}_{00}(m,r)=\frac{m^2G^2}{b^2}(-\frac{15}{4}+\frac{3r^2}{2b^2}+\frac{r^4}{4b^4}), \quad                  h^{(2)iso}_{\alpha\beta}(m,r)=\frac{m^2G^2}{b^2}\delta_{\alpha\beta}(\frac{15}{4}-\frac{3r^2}{b^2}+\frac{3r^4}{4b^4}).                                                                                                                                                                       \eqno(2.10)
$$
It is easy to verify that
$$
h_{\alpha\beta}^{(2)harm}(m,r)-h_{\alpha\beta}^{(2)iso}(m,r)=-\frac32\:{}^{(0)}\Lambda_{\alpha,\beta}+
\frac32\:{}^{(2)}\Lambda_{\alpha,\beta}-\frac12\:{}^{(4)}\Lambda_{\alpha,\beta}.                                                                                                                            \eqno(2.11)
$$
${}^{(n)}\Lambda_{\alpha,\beta}$ are gauges defined in Appendix, see (A19).

{\bf Perturbation approach to solving gravitational equation}\\
Now we turn  our attention to Einstein equation written in field theoretical form by Weinberg, see \S 6, Ch. 7
in [1].
$$
2R^{(1)}_{\alpha\beta}-\delta_{\alpha\beta}R^{(1)}=-16\pi G(T^{(1)}_{\alpha\beta}+t_{\alpha\beta}),\quad
R^{(1)}=R_{\alpha\alpha}^{(1)}-R_{00}^{(1)}=h^{(2)}_{,\alpha\alpha}-h^{(2)}_{\alpha\beta,\alpha\beta},  \eqno(2.12)
$$
$$
2R^{(1)}_{00}+R^{(1)}=-16\pi G(T^{(1)}_{00}+t_{00}),\quad  2R^{(1)}_{00}=\nabla^2h^{(2)}_{00}.    \eqno(2.13)
$$
Hear $R^{(1)}_{\alpha\beta}$ is linear in $h^{(2)}_{\alpha\beta}$ Ricci tensor:
$$
2R^{(1)}_{\alpha\beta}=h^{(2)}_{\alpha\beta,\gamma\gamma}-h^{(2)}_{\alpha\gamma,\gamma\beta}-
h^{(2)}_{\beta\gamma,\gamma\alpha}+h^{(2)}_{,\alpha\beta},\quad h^{(2)}=h^{(2)}_{\alpha\alpha}-h^{(2)}_{00}. \eqno(2.14)
$$
$T_{\alpha\beta}^{(1)}$ is the matter energy-momentum tensor:
$$
T_{\alpha\beta}^{(1)}=\delta_{\alpha\beta}p=\frac{m^2G^2}{8\pi Gb^4}\delta_{\alpha\beta}(3-\frac{3r^2}{b^2}),
$$
$$
 T_{00}^{(1)}=2\mu\phi=\frac{m^2G^2}{4\pi Gb^4}(\frac{3r^2}{b^2}-9).\quad\phi=\frac{mG}{2b}(\frac{r^2}{b^2}-3).                                                                                                                                    \eqno(2.15)
$$
Here $p$ and $\mu$ are pressure and (uniform) density in the ball of liquid.
The gravitational energy-momentum tensor $t_{ik}$ has the form
$$
t_{\alpha\beta}=\frac{1}{8\pi G}[\delta_{\alpha\beta}(4\phi\phi_{,\gamma\gamma}+3(\nabla\phi)^2)-4\phi\phi_{,\alpha\beta}-
$$
$$
2\phi_{,\alpha}\phi_{,\beta}]=
\frac{m^2G^2}{8\pi Gb^4}[\delta_{\alpha\beta}\left(-12+\frac{7r^2}{b^2}\right)-\frac{2x_{\alpha}x_{\beta}}{b^2}],\eqno(2.16)
 $$
$$
t_{00}=-\frac{3}{8\pi G}(\nabla\phi)^2-6\mu\phi=\frac{m^2G^2}{8\pi Gb^4}(54-\frac{21r^2}{b^2}). \eqno(2.17)
$$

From (2.15) and (2.17) we have
$$
T_{00}^{(1)} +t_{00}=\frac{m^2G^2}{8\pi Gb^4}(36-\frac{15r^2}{b^2}), \eqno(2.18)
$$
and from (2.15) and (2.16) we get
$$
T_{\alpha\beta}^{(1)}+t_{\alpha\beta}=\frac{m^2G^2}{8\pi Gb^4}[\delta_{\alpha\beta}\left(-9+\frac{4r^2}{b^2}\right)-\frac{2x_{\alpha}x_{\beta}}{b^2}],
                                                                                                 \eqno(2.19)
  $$

  From perturbation theory in first approximation we have
  $$
  h_{ik}^{(1)}(m',r)=-2\phi(m',r)\delta_{ik}, \quad \bar h_{ik}^{(1)}(m',r)=-4\phi(m',r)\delta_{i0\delta_{k0}},\quad \bar h_{ik}=
  h_{ik}-\frac12\eta_{ik}h.                                                                 \eqno(2.20)
  $$
  From perturbation theory in general relativity we know that in considered approximation $m=m'(1+3mG/b)$,
  see Duff [6], $\lambda^{Duff}=m'$. (It is somewhat surprising that $m>m'$.)

  By definition
  $$
   h_{ik}^{(1)}(m',r)+h_{ik}^{(2)}(m',r)=h_{ik}^{(1)}(m,r)+h_{ik}^{(2)}(m,r),\quad i,k=0,1,2,3.                                                                                                         \eqno(2.21)
  $$
  So, in general relativity we know $h^{(2)iso}_{ik}(m,r)$ from exact solution (2.10) and $h_{ik}^{(2)}(m',r)$ from (2.21). Indeed, from the first eq. (2.20)
and the last eq. (2.3) we have
$$
h_{00}^{(1)}(m',r)=h_{00}^{(1)}(m,r)+\frac{m^2G^2}{b^2}(-9+\frac{3r^2}{b^2}),
$$
$$
$$
$$ h_{\alpha\beta}^{(1)}(m',r)=h_{\alpha\beta}^{(1)}(m,r)+\frac{m^2G^2}{b^2}\delta_{\alpha\beta}(-9+\frac{3r^2}{b^2}).                                                                          \eqno(2.21a)
$$
From here and (2.21) we obtain
$$
h_{00}^{(2)}(m',r)=\frac{m^2G^2}{b^2}(\frac{21}{4}-\frac{3r^2}{2b^2}+\frac{r^4}{4b^4}), \quad h_{\alpha\beta}^{(2)iso}(m',r)=\frac{m^2G^2}{b^2}(\frac{51}{4}-\frac{6r^2}{b^2}+\frac{3r^4}{4b^4}).                                                                                              \eqno(2.21b)
$$

  Next we will consider how   $h_{ik}^{(2)}(m',r)$ is defined by  (2.12-13).
Its structure in isotropic coordinates should be of the form
$$
h_{00}^{(2)}=\frac{m^2G^2}{b^2}(a_0+a_2\frac{r^2}{b^2}+a_4\frac{r^4}{b^4}),\quad
h_{\alpha\beta}^{(2)iso}=\frac{m^2G^2}{b^2}(c_0+c_2\frac{r^2}{b^2}+c_4\frac{r^4}{b^4}).          \eqno(2.22)
$$
Ignoring $a_0$ and $c_0$, which cannot be determined from differential equations, and inserting (2.22) in (2.12-13) we find from (2.12)
$$
2(a_2-c_2)=9,\quad 2(a_4-c_4)=-1,                                                  \eqno(2.23)
$$
and from (2.13)
$$
c_2=-6,\quad c_4=\frac34.                                                          \eqno(2.24)
$$
Using (2.24) in (2.23) we get $a_2=-3/2, a_4=1/4$ i,e. we obtain $h_{00}^{(2)iso}(m'.r)$ up to additive constant $a_0, $, see the first eq. in (2.21b).  Other
$h_{\alpha\beta}^{(2)}(m',r) $ may differ by  gauge functions.
If we use in (2.23) $a_2=3/2, a_4=1/4$ from (2.9) we get $c_2=-3, c_4=3/4, i.e. $ $h_{\alpha\beta}^{(2)iso}(m,r)$ up to a constant, see the second eq. in (2.10).
 Thus, both $h_{ik}^{(2)}(m',r)$ and $h_{ik}^{(2)}(m,r)$ satisfy (2.12).

Now using the relations
$$
2R^{(1)}_{\alpha\beta}=\nabla^2h^{(2)}_{\alpha\beta}-\bar h^{(2)}_{\alpha\gamma,\gamma\beta}-\bar h^{(2)}_{\beta\gamma,\gamma\alpha},\quad R^{(1)}=R^{(1)}_{\alpha\alpha}-R^{(1)}_{00}=
\nabla^2h^{(2)}-  h^{(2)}_{\gamma\alpha,\gamma\alpha}=\nabla^2h^{(2)}-\bar h^{(2)}_{\gamma\alpha,\gamma\alpha}-\frac12 h_{,\alpha\alpha}=
$$
$$
\frac12\nabla^2h^{(2)}-\bar h^{(2)}_{\beta\gamma,\beta\gamma},\quad  h^{(2)}_{\alpha\gamma,\alpha\gamma} =\bar h^{(2)}_{\alpha\gamma,\alpha\gamma}+\frac12\delta_{\alpha\gamma}h_{,\alpha\gamma},                                                                                                            \eqno(2.14a)
$$
we rewrite (2.12) in the form
$$
2R^{(1)}_{\alpha\beta}-\delta_{\alpha\beta}R^{(1)}=\nabla^2\bar h_{\alpha\beta}^{(2)}-
 -\bar h_{\alpha\gamma,\gamma\beta}^{(2)}-\bar h_{\beta\gamma,\gamma\alpha}^{(2)}+\delta_{\alpha\beta}\bar h_{\gamma\sigma,\gamma\sigma}^{(2)}
=-16\pi G(T^{(1)}_{\alpha\beta}+t_{\alpha\beta}).                                              \eqno(2.12a)
$$
This is a differential form of Schwinger's eq.(17.6) (in our notation). We remind that $\partial^2D_+(x-x')=-\delta_{\alpha\beta}(x-x')$ and we consider the static case. Dropping the gauge degrees of freedom in (2.12a) (i.e. using Hilbert gauge $\bar h^{(2)}_{\alpha\beta,\alpha}=0$) we write the integral form as follows
$$
\bar h_{\alpha\beta}^{(2)}(m',r)=16\pi G\int\frac{d^3x'}{4\pi}\frac{1}{|\vec x-\vec x'|}[T^{(1)}_{\alpha\beta}+
t_{\alpha\beta}].                                                          \eqno(2.12b)
$$
So Schwinger's formula (17.6) in Ch.3 in [2] holds also in our nonlinear case without modification.

{\bf Theory of sources approach}\\

The $h^{(2)}_{00}(m',r)$ is given by Schwinger, see eq. (17.4),  \S 17, Ch.3 in [2]. In our notation it has the form
$$
h^{(2)}_{00}(m',r)=16\pi G\int d^4x'D_+(x-x')[\bar T^{(1)}_{00}(x')+\bar t_{00}(x')],\quad \bar T_{00}^{(1)}= \frac32p+\mu\phi,                                                                                                                                                                \eqno(2.25)
$$
$$
 T^{(1)}=3p=\frac{m^2G^2}{8\pi G}(9-\frac{9r^2}{b^2}),\quad \bar t_{00}=t_{00}+\frac{1}{2}t=\frac{m^2G^2}{8\pi G}(9-\frac{r^2}{b^2}),
 $$
$$
 t=t_{\alpha\alpha}-t_{00},\quad 16\pi G(\bar T_{00}^{(1)}(x)+\bar t_{00}(x))=\frac{m^2G^2}{b^2}(9-\frac{5r^2}{b^2}).\eqno(2.26)
$$
We note that $\bar T_{00}^{(1)}$ is calculated from the relation
$$
\bar T_{00}=T_{00}^{(1)}-\frac12g_{00}T=\frac12\mu+\frac32p+\mu\phi=\bar T_{00}^{(0)}+\bar T_{00}^{(1)}.
$$
Using (A5) and (A6) we easily get from (2.25)
$$
h^{(2)}_{00}(m',r)=\frac{m^2G^2}{b^2}(\frac{21}{4}-\frac{3r^2}{2b^2}+\frac{r^2}{4b^2}). \eqno(2.27)
$$

Now Schwinger's formula (17.4) for $h^{(2)}_{\alpha\beta}(m',r)$  needs a correction:
$$
h^{(2)}_{\alpha\beta}(m',r)=16\pi G\int d^4x'D_+(x-x')[\bar T^{(1)cor}_{\alpha\beta}(x')+\bar t_{\alpha\beta}(x')], \eqno(2.28)
$$
$$
\bar T_{\alpha\beta}^{(1)cor}= T_{\alpha\beta}^{(1)}-\frac12\delta_{\alpha\beta}T^{(1)cor},\quad  T^{(1)cor}_{\alpha\beta}=
\eta^{ik}T^{(1)}_{ik}=3p-2\mu\phi= \frac{m^2G^2}{8\pi Gb^4}(27-\frac{15r^2}{b^2}),                            \eqno(2.29)
$$
$$
\bar t_{\alpha\beta}= t_{\alpha\beta}-\frac12\delta_{\alpha\beta}t,\quad t=t_{\alpha\alpha}-t_{00},\quad \bar T^{(1)cor}_{\alpha\beta}+
\bar t_{\alpha\beta}=\frac{m^2G^2}{16\pi Gb^4}[\delta_{\alpha\beta}(45-\frac{17r^2}{b^2})-\frac{4x_{\alpha}x_{\beta}}{b^2}],
                                                                                                                                                                                                                            \eqno(2.30) 
 $$
$$
\bar t_{\alpha\beta}=\frac{m^2G^2}{16\pi Gb^4}[\delta_{\alpha\beta}(66-\frac{26r^2}{b^2})-\frac{4x_{\alpha}x_{\beta}}{b^2}].          
$$
Then using (A5-6), (A12), (A17) we get
$$
h^{(2)}_{\alpha\beta}(m',r)=\frac{m^2G^2}{b^2}[\delta_{\alpha\beta}(\frac{61}{4}-\frac{69r^2}{10b^2}+\frac{23r^4}{28b^4})-
\frac{9x_{\alpha}x_{\beta}}{5b^2}+\frac{2r^2x_{\alpha}x_{\beta}}{7b^4}].                                  \eqno(2.31)
$$
Putting here $r=b$ we obtain 
$$
h^{(2)}_{\alpha\beta}(m',b)=\frac{m^2G^2}{b^2}[\frac{321}{35}\delta_{\alpha\beta}-\frac{53x_{\alpha}x_{\beta}}{35}]
                                                                                                                               \eqno(2.31a)
$$
Next we want to calculate barred quantity. We use Schwinger's eq.(17.6) putting his gauge vector $\xi$ to zero.
In our notation we have
$$
\bar h_{00}^{(2)}(m',r)=16\pi G\int d^4x'D_+(x-x')[T_{00}^{(1)}(x')+t_{00}(x')],                           \eqno(2.32)
$$

$$
\bar h_{\alpha\beta}^{(2)}(m',r)=16\pi G\int d^4x'D_+(x-x')[T_{\alpha\beta}^{(1)}(x')+t_{\alpha\beta}(x')], \eqno(2.33)
$$
The sources are defined in (2.18), (2.19).

Using (A5) and (A6) we find from (2.32)
$$
\bar h_{00}^{(2)}(m',r)=\frac{m^2G^2}{b^2}(\frac{51}{2}-\frac{12r^2}{b^2}+\frac{3r^4}{2b^4}).  \eqno(2.34)
$$
Similarly to (2.21) we have
 $$
  \bar h_{00}^{(1)}(m',r)+\bar h_{00}^{(2)}(m',r)=\bar h_{00}^{(1)}(m,r)+\bar h_{00}^{(2)}(m,r).                 \eqno(2.21a)
  $$
  $$
  \bar h_{00}^{(2)}(m,r)=\frac{m^2G^2}{b^2}(\frac{15}{2}-\frac{6r^2}{b^2}+\frac{3r^4}{2b^4}). \eqno(2.35)
  $$
  $$
  \bar h_{00}^{(1)}(m',r)=-4\phi(m',r)=\frac{m'G}{b}(6-\frac{2r^2}{b^2})=\bar h_{00}^{(1)}(m,r)+\frac{m^2G^2}{b^2}(-18+\frac{6r^2}{b^2}),                                                                                                                    \eqno(2.36)
  $$

 From (2.33) using (A12) and (A17) we obtain
$$
\bar h^{(2)}_{\alpha\beta}(m',r)=\frac{m^2G^2}{b^2}[\delta_{\alpha\beta}(-5+\frac{18r^2}{5b^2}-\frac{3r^4}{7b^4})-
\frac{9x_{\alpha}x_{\beta}}{5b^2}+\frac{2r^2x_{\alpha}x_{\beta}}{7b^4}].                                  \eqno(2.37)
$$

From (2.37) and (2.34) we have
$$
\bar h^{(2)}(m',r)=\bar h^{(2)}_{\alpha\alpha}(m',r)-\bar h_{00}^{(2)}(m',r)=\frac{m^2G^2}{b^2}(-\frac{81}{2}+\frac{21r^2}{b^2}-\frac{5r^4}{2b^4}).           \eqno(2.38)
$$
It is easy to check that as it should be
$$
-\bar h^{(2)}(m',r)= h^{(2)}(m',r)= h^{(2)}_{\alpha\alpha}(m',r)- h^{(2)}_{00}(m',r). \eqno(2.39)                                                           $$

We note that harmonic $h_{\alpha\beta}^{(2)har}$ differs from privileged system $h_{\alpha\beta}^{(2)}$
only by a gauge function:
$$
h_{\alpha\beta}^{(2)har}(m,r)-h_{\alpha\beta}^{(2)}(m,r)=h_{\alpha\beta}^{(2)har}(m',r)-h_{\alpha\beta}^{(2)}(m',r)=
-4\:{}^{(0)}\Lambda_{\alpha,\beta}+\frac{12}{5}\:{}^{(2)}\Lambda_{\alpha,\beta}-\frac{4}{7}\:{}^{(4)}\Lambda_{\alpha,\beta}
$$
$$
=\frac{m^2G^2}{b^2}[\delta_{\alpha\beta}(-4+\frac{12r^2}{5b^2}-\frac{4r^4}{7b^4})+\frac{24x_{\alpha}x_{\beta}}{5b^2}-
\frac{16r^2x_{\alpha}x_{\beta}}{7b^4}].                                                             \eqno(2.40)
$$

As we have seen earlier $\bar h^{(2)}_{\alpha\beta}(m',r)$ can be calculated by Schwinger formula without modification. The return to $ h^{(2)}_{\alpha\beta}(m',r)$ can be made with the help of formula
$$
 h^{(2)}_{\alpha\beta}(m',r)=\bar h^{(2)}_{\alpha\beta}(m',r)-\frac{1}{2}\delta_{\alpha\beta}\bar h^{(2)}(m',r), \eqno(2.41)
$$
see eqs.(2.37) and (2.38).

 Similarly
$$
 h^{(2)}_{00}(m',r)=\bar h^{(2)}_{00}(m',r)+\frac{1}{2}\bar h^{(2)}(m',r). \eqno(2.42)
$$

 \section{Exterior solution}
 The linear approximation is well known:
 $$
 h^{(1)}_{ik}(m',r)=-2\delta_{ik}\phi(m',r),\quad \phi(m',r)=-\frac{m'G}{r},\quad \bar  h^{(1)}_{ik}(m',r)=-4\delta_{i0}\delta_{k0}\phi(m',r).                                    \eqno(3.1)
 $$
 So we have only to obtain $h^{(2)}_{ik}(m',r)$. Contribution to $h^{(2)}_{\alpha\beta}(m',r)$ from  $r'<b$ is
$$
16\pi G\int_{r'<b<r}\frac{d^3x'}{4\pi}\frac{1}{|\vec x-\vec x'|}[\bar T^{(1)cor}_{\alpha\beta}(x')+\bar t_{\alpha\beta}(x')]=
m^2G^2[\frac{34\delta_{\alpha\beta}}{3rb}+\frac{4b}{35}\left(\frac{\delta_{\alpha\beta}}{3r^3}-
\frac{x_{\alpha}x_{\beta}}{r^5}\right)].                                                         \eqno(3.2)
$$
The source in the integrand is given in eq. (2.30).

 Contribution to $h^{(2)}_{\alpha\beta}(m',r)$ from  $r'>b$ is
$$
16\pi G\int_{b<r',r}\frac{d^3x'}{4\pi}\frac{1}{|\vec x-\vec x'|}\bar t_{\alpha\beta}(x')=
m^2G^2[-\frac{16\delta_{\alpha\beta}}{3rb}+
(\frac{5\delta_{\alpha\beta}}{r^2}-
\frac{7x_{\alpha}x_{\beta}}{r^4})+\frac{28b}{5}(\frac{x_{\alpha}x_{\beta}}{r^5}-\frac{\delta_{\alpha\beta}}{3r^3})], \eqno(3.3)
$$
$$
16\pi G\bar t_{\alpha\beta}=m^2G^2(\frac{4\delta_{\alpha\beta}}{r^4}-\frac{28x_{\alpha}x_{\beta}}{r^6}).
$$
Adding (3.2) and (3.3) we obtain
$$
h^{(2)}_{\alpha\beta}(m',r)=m^2G^2[\frac{6\delta_{\alpha\beta}}{rb}+ (\frac{5\delta_{\alpha\beta}}{r^2}-
\frac{7x_{\alpha}x_{\beta}}{r^4})+\frac{192b}{35}(\frac{x_{\alpha}x_{\beta}}{r^5}-\frac{\delta_{\alpha\beta}}{3r^3})].
                                                                                                                      \eqno(3.4)
$$
The first term in the r.h.s. of (3.4) can be written in the form $\frac{m'mG^26\delta_{\alpha\beta}}{rb}$. Joining it to $h^{(1)}_{\alpha\beta}(m',r)=2m'G/r$ we get $h^{(1)}_{\alpha\beta}(m,r)=2mG/r, m=m'(1+\frac{3mG}{b})$, and $h^{(2)}_{\alpha\beta}(m',r)$ without first term in the r.h.s. of (3.4) is by definition $h^{(2)}_{\alpha\beta}(m,r)$.

Putting $r=b$ in (3.4) we get the relation (2.31a). The term (proportional to $b$) having the gauge form plays an important role in forming the continuity at $r=b$.
 
Similarly we deal with $h^{(2)}_{00}(m',r)$.  Contribution from $r'<b$ is
$$
16\pi G\int_{r'<b<r}\frac{d^3x'}{4\pi}\frac{1}{|\vec x-\vec x'|}[\bar T^{(1)}_{00}(x')+\bar t_{00}(x')]=
m^2G^2\frac{2}{rb}.                                                                                  \eqno(3.5)
$$
The source is given in (2.26).

Contribution from $r'>b$ is
$$
16\pi G\int_{b<r',r}\frac{d^3x'}{4\pi}\frac{1}{|\vec x-\vec x'|}\bar t_{00}(x')=
m^2G^2(\frac{4}{rb}-\frac{2}{r^2}),\quad 16\pi G\bar t_{00}(x)=\frac{4m^2G^2}{r^4}.\eqno(3.6)
$$
Adding (3.5) and (3.6) we get
$$
h^{(2)}_{00}(m',r)=m^2G^2(\frac{6}{rb}-\frac{2}{r^2}), \quad \mbox{and hence}
\quad h^{(2)}_{00}(m,r)=-m^2G^2\frac{2}{r^2}.                                         \eqno(3.7)
$$

We remind here that according (3.1)
$$
h^{(1)}_{00}(m',r)=-2\phi(m',r)=\frac{2m'G}{r}=\frac{2mG}{r}(1-\frac{3mG}{b}),\mbox{and hence}\quad h^{(1)}_{00}(m,r)=\frac{2mG}{r}.
 \eqno(3.8)
$$

Now from (3.4) we have
$$
h^{(2)}_{\alpha\alpha}(m',r)=m^2G^2\left(\frac{18}{rb}+\frac{8}{r^2}\right). \eqno(3.9).
$$
From  (3.7) and (3.9) we have
$$
h^{(2)}(m',r)=h^{(2)}_{\alpha\alpha}(m',r)-h^{(2)}_{00}(m',r)=m^2G^2\left(\frac{12}{rb}+\frac{10}{r^2}\right),
\mbox{and so}\: h^{(2)}(m,r)= \frac{10 m^2G^2}{r^2}.                                        \eqno(3.10)
$$

In the same manner we obtain
$$
\bar h^{(2)}_{00}(m',r)=m^2G^2\left(\frac{12}{rb}+\frac{3}{r^2}\right),\quad \bar h^{(2)}_{00}(m,r)=m^2G^2\frac{3}{r^2}, \eqno(3.11)
$$
$$
\bar h^{(2)}_{\alpha\beta}(m',r)=m^2G^2[-\frac{7x_{\alpha}x_{\beta}}{r^4}+
\frac{192b}{35}\left(\frac{x_{\alpha}x_{\beta}}{r^5}
-\frac{\delta_{\alpha\beta}}{3r^3}\right)]=\bar h^{(2)}_{\alpha\beta}(m,r).\eqno(3.12)
$$
The last equation is in agreement with the fact that $\bar h^{(1)}_{\alpha\beta}(m,r)=0$, see the last  eq. in (3.1).
And finally
$$
\bar h^{(2)}(m,r)=\bar h^{(2)}_{\alpha\alpha}(m,r)-\bar h^{(2)}_{00}(m,r)=-\frac{10m^2G^2}{r^2}=-h^{(2)}(m,r).\eqno(3.13)
$$
The last equation is in agreement with the last equation in (3.10).
\section{Conclusion}
No modification is needed in linear source theory if we first calculate $\bar h^{(2)}_{ik}$ and then go to $h^{(2)}_{ik}$.

Generally speaking gauge transformations are unobservable as in classical electrodynamics, but in some cases they can be interpreted as coordinate transformations. This is the case for harmonic and isotropic coordinate systems.

Quite specific term (having the form of a gauge function) enters the privileged exterior metric. I think that it should be observable.
I expect that when gravity is turned on (i.e. we assemble adiabatically the ball of matter) we find ourself in the privileged coordinate system.
\section{Appendix}
Here I give some formulas needed for calculating the  metric. The propagator is denoted
as
$$
D_+(x)=\left(\frac{1}{2\pi}\right)^2\frac{i}{x^2+i\epsilon}.                    \eqno(A1)
$$
$$
\int dtD(\vec x,\vec x',t)=\frac{1}{(2\pi)^2}\int_{-\infty}^{\infty}\frac{1}
{|\vec x-\vec x'|^2-t^2+i\epsilon}=\frac{1}{4\pi|\vec x-\vec x'|}.                                 \eqno(A2)
$$
In the following I use the notation: at the foot of an integral I indicate the region of integration and , if necessary, also the assumed position of $r$ as in (A12) and other formulas.
$$
\int_{r'<b}\frac{d^3x'}{4\pi}\frac{1}{|\vec x-\vec
x'|}x'_{\alpha}x'_{\beta}=x_{\alpha}x_{\beta}A(r)+\delta_{\alpha\beta}B(r).                  \eqno(A3)
$$
The functions $A(r)$ and $B(r)$ can be obtained in the following way. First we put
$\alpha=\beta$:
$$
\int_{r'<b}\frac{d^3x'}{4\pi}\frac{1}{|\vec x-\vec x'|}r'^2=r^2A(r)+3B(r).                 \eqno(A4)
 $$
The integral on the l.h.s. is easy to evaluate:
$$
\int_{r'<b}\frac{d^3x'}{4\pi}\frac{1}{|\vec x-\vec x'|}r'^2=\left\{\begin{array}{cc}
                                                                     \frac 14b^4-\frac{1}{20}r^4,\ & \quad\  r<b\\
                                                                     \frac{b^5}{5r}, & \quad\ r>b.
\end{array}\right.                                                                              \eqno(A5)
$$
Similarly we obtain the expression needed in another place

$$
\int_{r'<b}\frac{d^3x'}{4\pi}\frac{1}{|\vec x-\vec x'|}=\left\{\begin{array}{cc}
\frac12b^2-\frac{1}{6}r^2,&\quad r<b\\
\frac{b^3}{3r}, & \quad r>b,
\end{array}\right.
                                                                                      \eqno(A6)
$$
In (A5) and (6) we used the relation
$$
\int_{-1}^1\frac{dt}{\sqrt{r^2+r'^2-2rr't}}=
\left\{\begin{array}{cc}\frac{2}{r},&\quad r'<r\\\frac{2}{r'},& \quad r'>r,\end{array}\right.
                                                                                       \eqno(A7)
$$
Similarly we obtain
$$
\int_1^1\frac{t^2dt}{\sqrt{r^2+r'^2-2rr't}}=\left\{\begin{array}{cc}
\frac{2}{3r}+\frac{4r'^2}{15r^3},&\quad r'<r\\
\frac{2}{3r'}+ \frac{4r^2}{15r'^3},&\quad r'>r,
\end{array}\right.
                                                                                      \eqno(A8)
$$
Returning to (A3) we put $ \alpha=\beta=3 $ and assuming that $\vec x$ is directed along 3-axes we get
$$
\int_{r'<b}\frac{d^3x'}{4\pi}\frac{1}{|\vec x-\vec x'|}r'^2t^2=r^2A(r)+B(r).
                                                                                    \eqno(A9)
$$
Evaluating the integral in the l.h.s. of (A9) with the help of (A8)we find for $r<b$
$$
r^2A(r)+B(r)=\frac{b^4}{12}+
\frac{b^2r^2}{15}-\frac{9r^4}{140}.                                              \eqno(A10)
$$
From (A4)and (A5) we get for $r<b$

$$
\frac{b^4}{4}-\frac{r^4}
{20}=r^2A(r)+3B(r).                                                       \eqno(A11)
$$
From (A10) and (A11) we get for $r<b$
$$
A(r)=\frac{b^2}{10}-\frac{r^2}{14}, \quad B(r)=\frac{b^4}{3\cdot4}-
\frac{b^2r^2}{2\cdot3\cdot5}+\frac{r^4}{4\cdot5\cdot7}.                            \eqno(A11^1)
$$
Finally from (A3) and $(A11^1)$ we obtain
$$
\int_{r,r'<b}\frac{d^3x'}{4\pi}\frac{1}{|\vec x - \vec x'|}x'_{\alpha} x'_{\beta}=
\frac{r^4\delta_{\alpha\beta}}{140}-
\frac{r^2x_{\alpha}x_{\beta}}{14}+\frac{b^2}{10}(x_{\alpha}x_{\beta}-
\frac{r^2}{3}\delta_{\alpha\beta})+
\frac{b^4}{12}\delta_{\alpha\beta}.                                                    \eqno(A12)
$$
Only first two terms in the r.h.s. can be obtained from differential equation.
Term proportional $b^2$ is due to the boundary $r'=b$. Indeed, if we use in the
integrand in (A12) the relation
$$
x'_{\alpha}x'_{\beta}=-\nabla^2(\frac{r'^4\delta_{\alpha\beta}}{140}-
\frac{r'^2x'_{\alpha}x'_{\beta}}{14}),
                                                                                \eqno(A13)
$$

integrate twice by parts and neglect the contribution from the boundary $r'=b$ we
get the first two terms in the r.h.s. of (A12). Here use is made of the relation
$$
\nabla^2\frac{1}{|\vec x-\vec x' |}= - 4\pi\delta(\vec x -\vec x').
                                                                                        \eqno(A14)
$$
 We note also that  the term proportional to $b^2$ in (A12) has no source because $\nabla^2(x_{\alpha}x_{\beta}-\frac13r^2\delta_{\alpha\beta})=0$ and it is not a gauge function.

The following formulas are obtained similarly to the above ones.
$$
\int_{b<r',r}\frac{d^3x'}{4\pi}\frac{1}{|\vec x-\vec
x'|}
\frac{x'_{\alpha}x'_{\beta}}{r'^6}=\delta_{\alpha\beta}\frac{1}{3rb}+x_{\alpha}x_{\beta}\frac{1}
{4r^4} -\delta_{\alpha\beta}\frac{1}{4r^4}+
\frac{b}{5}(\frac{\delta_{\alpha\beta}}{3r^3}-\frac{x_{\alpha}x_{\beta}}{r^5}).
                                                                                             \eqno(A15)
$$
$$
\int_{r'<b<r}\frac{d^3x'}{4\pi}\frac{1}{|\vec x-\vec
x'|}\frac{x'_{\alpha}x'_{\beta}}{b^6}= \frac{\delta_{\alpha\beta}}{15rb}+
\frac{b}{35}(\frac{x_{\alpha}x_{\beta}}{r^5}-\frac{\delta_{\alpha\beta}}
{3r^3}).                                                                                      \eqno(A16)
$$

$$
\int_{r<b<r'}\frac{d^3x'}{4\pi}\frac{1}{|\vec x-\vec
x'|}\frac{x'_{\alpha}x'_{\beta}}{r'^6}=\frac{\delta_{\alpha\beta}}{6b^2}+\frac{1}{20b^4}(x_{\alpha}x_{\beta}-
\frac{\delta_{\alpha\beta}r^2}{3}).
                                                                                                                  \eqno(A17)
$$
 Putting $\alpha=\beta$ in (A17) and (A15) we find                                                                                          $$
\int_{b<r'}\frac{d^3x'}{4\pi}\frac{1}{|\vec x-\vec x'|}\frac{1}{r'^4}=\left\{\begin{array}{cc}
                                                                     \frac {1}{2b^2},\ & \quad\  r<b\\
                                                                     \frac{1}{rb}-\frac{1}{2b^2}, & \quad\ r>b.
\end{array}\right.                                                                              \eqno(A18)
$$

If we act by $\nabla^2$ on (A15) or (A 17) we get zeros; in the l.h.s. because the argument of $\delta$-function
is not zero, see (A14)and low limits of integrals.

Finally I give the family of gauge functions:
$$
^{(n)}\Lambda_{\alpha,\beta}=\frac{m^2G^2}{b^2}(x_{\alpha}\frac{r^n}{b^n})_{,\beta}=
\frac{m^2G^2}{b^2}\left(\delta_{\alpha,\beta}\frac{r^n}{b^n}+n\frac{r^{n-2}x_{\alpha}x_{\beta}}{b^n}\right),\quad n=0, \mp1,\mp2,\cdots.                                                                            \eqno(A19)
$$
and fictitious sources:
$$
\nabla^2\:{}^{(n)}\Lambda_{\alpha,\beta}=
n(n+3)\frac{m^2G^2}{b^{n+2}}[\delta_{\alpha,\beta}r^{n-2}+
(n-2)r^{n-4}x_{\alpha}x_{\beta}].                                         \eqno(A20)
$$
We note that for $n=-3$ the fictitious source is absent: $
\nabla^2\:{}^{(-3)}\Lambda_{\alpha,\beta}=0
$
and besides
$$
{}^{(-3)}\Lambda_{\alpha,\alpha}= {}^{(-3)}\Lambda_{\alpha,\beta \beta}=0.                 \eqno(A21)
$$
The last two equations mean that Hilbert coordinate condition is satisfied.
\section*{References}
1. Weinberg S. Gravitation and Cosmology. New York 1972.\\
2. Schwinger J. Particles, sources. and fields. Addison- Wesley, 1970.\\
3. Nikishov A.I. Physics of Particles and Nuclei,Vol. 37, No3, pp.776=784, (2006).\\
4. Synge J.I., Relativity. The general theory, Amsterdam (1960).\\
5. Rosen N. Ann. Phys. (N.Y.), {\bf 63}, 127, (1970).\\
6. Duff M.J. Phys.Rev. D, 7, 2317, (1973).\\
\end{document}